# Improving the nonlinear performance of the HEPS baseline design with genetic algorithm*

JIAO Yi(焦毅)[1)]

Key Laboratory of Particle Acceleration Physics and Technology, Institute of High Energy Physics,

Chinese Academy of Sciences, Beijing 100049, China

**Abstract:** The baseline design for the High Energy Photon Source has been proposed, with an emittance of 60 pm·rad within a circumference of 1.3 kilometers. Nevertheless, the nonlinear performance of the design needs further improvements to enlarge both the dynamic aperture and the momentum acceptance. In this study, genetic optimization of the linear optics is performed, so as to find all the possible solutions with weaker sextupoles and hence weaker nonlinearities, while keeping the emittance at the same level as the baseline design. These obtained solutions enable us to explore the dependence of nonlinear dynamics on the working point. The result indicates that with the same layout, it is feasible to obtain much better nonlinear performance with a delicate tuning of the magnetic field strengths and a wise choice of the working point.

## 1 Introduction

Along with the continuous advance in accelerator technology and unceasing pursuit of higher quality photon flux, worldwide efforts [1] are being made to push the brightness and coherence beyond the existing third generation light sources (3GLSs), by reducing the emittance to approach the diffraction-limit for the range of X-ray wavelengths of interest to the scientific community. Such new-generation rings are usually called diffraction-limited storage rings (DLSRs). Different form the 3GLSs which typically consist of double-bend achromats or triple-bend achromats, in a DLSR design multi-bend achromat (MBA) lattice with compact layout as well as strong focusing is usually adopted to achieve an ultralow emittance within a reasonable circumference [2-4]. However, along with the decrease in emittance, nonlinear dynamics grows to be a great challenge to the global ring performance. The strong focusing creates large chromaticities and small dispersions. This, in turn, requires very strong sextupoles for chromatic correction. The accompanied extremely strong nonlinearities can limit the dynamic aperture (DA) to the order of a few millimeters and cause the momentum acceptance (MA) smaller than 3% (typical required value in 3GLSs), leading to a short beam lifetime and a poor injection efficiency. Thus, how to control these strong nonlinear effects and obtain adequate DA and MA is a very important issue in a DLSR design.

A kilometer-scale DLSR with the beam energy of 5 to 6 GeV, named High Energy Photon Source (HEPS), was proposed to be built in Beijing [5]. Extensive efforts have been made on the lattice design and relevant studies for this project [6-14]. Recently the 'baseline design' of the HEPS was developed [15] based on the concept of 'hybrid MBA' [16].The ring consists of 48 identical hybrid

* Supported by NSFC (11475202, 11405187) and Youth Innovation Promotion Association CAS (2015009)

1) jiaoyi@ihep.ac.cn

seven-bend achromats (7BAs), with a horizontal natural emittance of 60 pm·rad and a circumference of 1.3 kilometers. The layout and the optical functions of a single 7BA are shown in Fig. 1, and the main parameters of the ring are listed in Table 1.

This design adopts high-gradient (up to 80 T/m) quadrupoles and three dipoles with horizontal defocusing gradients in the middle two unit cells of the 7BA, to create a compact layout and an ultralow emittance. Moreover, it utilizes four outer dipoles with longitudinal gradients to generate two dispersion bumps, for a more efficient chromatic correction than available in a normal MBA. Moreover, a $-\boldsymbol{I}$ transportation between each pair of sextupoles, with phase advance at or close to $(2n+1)\pi$ ($n$ is integer) in both $x$ and $y$ planes, is designed to cancel most of the nonlinearities induced by sextupoles,. To reserve as much space as possible to accommodate various hardware systems, only four families of multipoles are placed in the dispersion bumps (see Fig. 1), with two families of sextupoles to correct the chromaticities and the other two families (one is sextupole and the other is octupole) to control the nonlinear driving terms. Due to the small number of free variables, we were able to do a grid scan of the DA and MA with respect to the multipole strengths in a reasonable time. However, it was found very difficult, if not impossible, to simultaneously optimize the DA and MA, even with the adoption of the dispersion bumps and the local cancellation scheme. The compromised solution predicts an 'effective' DA of 2.2 mm in the vertical (injection) plane and a MA of 2.4%. Since magnetic and misalignment errors were not considered during the particle tracking, the effective DA which counts only the stable motions with small tune variations (i.e., with tunes in the range of [0, 0.5]), rather than the DA defined in a conventional sense (i.e., the area with all survived particles after tracking over typically a few thousand turns), is used here to give a more accurate estimation of the actual ring acceptance. For more details about the HEPS baseline design, one can refer to Ref. [15].

To achieve a robust machine performance for the HEPS design, obviously it is necessary to further enlarge the DA and MA. Generally speaking, to achieve larger DA and MA, it requires a less aggressive transverse focusing and weaker chromatic sextupoles which, however, are usually in conflict with the pursuing of a minimum emittance for high brightness in a DLSR. Furthermore, the nonlinear optimization and linear matching are always coupled. In other words, the nonlinear performance relies not only on the strengths of the nonlinear elements (multipoles), but also on the choice of the linear optical parameters. On the other hand, as the main knobs, the magnetic field strengths cannot be varied arbitrarily, but within specific ranges limited by the available magnet technology. For such a nonlinear problem with conflicting objectives and many variables, the evolutionary genetic algorithm is suitable to find solutions with all possible tradeoffs between different objectives. For the optimization of the ring performance, one can directly use the DA and/or MA (from time-consuming numerical tracking) as the optimization objectives [see, e.g., 17, 18]. This approach is straightforward, but usually costs a lot of computing time and resources.

Instead, in this study we constrain the genetic optimization in the linear optics regime, where there exist definite relations between the variables (e.g., quadrupole strengths) and the objectives (e.g., the emittance and the optical functions) and the evaluation of the objectives is much faster. Before carrying out the subsequent nonlinear optimization, lots of solutions with unfavorable optics can be filtered according to prior knowledge and experience. Furthermore, from the post-optimization based on the survived solutions, one can also get an understanding of the dependence of nonlinear dynamics on specific optical parameters. As will be shown in Sec. 2, this approach enables us to significantly improve the nonlinear performance of the HEPS baseline design. At last, conclusions will be given in Sec. 3.

## 2 Improving the nonlinear performance with genetic algorithm

As mentioned, in our study we first perform a global scan of the linear optics with an evolutionary genetic method, called non-dominated sorting genetic algorithm II (NSGA-II [19]). This algorithm mimics natural selection from an initial random population with $N$ individuals, evolves generation by generation, until reaching a generation with desired convergence to the so-called Pareto optimal front with solutions showing all the possible tradeoffs between the different objectives.

The first step is to determine the variables, objectives, and constraints. To keep the lattice structure the same in the optimization, only the gradients of the separate-function quadrupoles ($Qi$, with $i$ = 1 to 8, see Fig. 1) are varied. The goal is to simultaneously minimize the emittance (the main indicator of linear performance) and maximize the dispersion at the sextupole, SF, nearly the center of the dispersion bump. Note that the value of the dispersion is closely related to the chromatic sextupole strengths and hence the difficulty of the DA and MA optimization. Thus, the dispersion at the sextupole is a good indicator of the nonlinear performance. In addition, to ensure satisfactory linear performances of the solutions, several constraints are considered, which are

(1) the achromatic condition, with zero dispersion functions ($D_x$, $D'_x$) at the long straight section for insertion device (ID) or injection;
(2) –***I*** transportation between sextupole pairs, with both the horizontal and vertical phase advance between the pair of sextupole, SF, exactly at or close to $(2n + 1)\pi$, where $n$ is integer;
(3) maximum gradients of 52 T/m for nominal quadrupoles ($Qi$ with $i$ = 1 to 6) and of 80 T/m for high-gradient quadrupoles ($Qi$ with $i$ = 7 to 8);
(4) the fractional part of the working point within the range of [0, 0.5], which is favorable against the resistive wall instability;
(5) reasonably low beta functions at the ID section for high brightness, with 1.5 m < $\beta_y$ < 4 m and 1.5 m < $\beta_x$ < 15 m;
(6) reasonably low natural chromaticities, with $|\xi_x| < 4.5$ and $|\xi_y| < 3.5$ in each hybrid 7BA.

To distinguish the desirable solutions from those which do not satisfy the constraints, a series of weight factors $\lambda_i$ (with $i$ = 1 to 6) are defined to measure the degree of the violation of the constraints. For example, $\lambda_1$ is given by

$$\lambda_1 = \begin{cases} 1, & if \quad \sqrt{D_x^2[m^2] + D_x'^2} \le 10^{-4}, \\ \sqrt{D_x^2[m^2] + D_x'^2}\,/10^{-4}, & if \quad \sqrt{D_x^2[m^2] + D_x'^2} > 10^{-4}. \end{cases} \quad (1)$$

The calculated emittance and the dispersion at the sextupole, multiplied by these weight factors, form the two optimization objectives.

Afterwards, the optimization can be preceded immediately. In this step, a random population of $N$ = 2000 individuals is generated and evolved generation by generation, until almost all of the solutions converge to the regions where the constraints are satisfied. The results after evolution of 120 generations are shown in Fig. 2, where the values of the two objectives corresponding to the baseline design is also plotted (as a black dot) for comparison. It appears feasible to obtain many candidate solutions with emittance of about 60 pm.rad and dispersion (at the sextupole) of above 5.85 cm.

Among the obtained solutions, only those with emittances below 70 pm.rad and with $\lambda_i = 1$ ($i = 1$ to 6) are kept and then re-evaluated to get the other optical parameters, including the optical functions, working point, natural chromaticities, and the required chromatic sextupole strengths. And then, we can visualize these results in different sub-optical parameter spaces, e.g., in the tune space and the sextupole strength space (see Fig. 3). It shows that the solutions, although not evenly distributed, spread over the whole tune space; and the required sextupole strengths are much smaller than those in the baseline design which are (–250, 337) $m^{-3}$. It is worth mentioning that for the sake of the comparison of the sextupole strengths between different solutions, it is assumed that the sextupoles have identical lengths of 0.2 m and are grouped in only two families such that there is a unique solution to correct the chromaticities to specific positive values ([0.5, 0.5] in this study).

As is known, the working point, which reflects the phase advance between multipoles, is closely related to the strengths of the nonlinear driving terms and hence the nonlinear performance of the ring [20]. Thus, a tune space survey within the range of [0.1, 0.4] in both $x$ and $y$ planes is performed, through a grid scan of the DA and MA with respect to the multipole strengths for each specific working point. Note that in this case, the sextupoles are divided into three families again, with the same lengths as those in the baseline design (see Table 1). From statistics, the dependence of the available effective vertical DA on the working point, at different levels of MA, can be explored, as shown in Fig. 4. One can see the largest DA can be obtained around (0.15, 0.15) in the tune space. By looking insight the dynamics with frequency map analysis [21], we reach a set of magnetic field strengths, promising an effective vertical DA of ~3.5 mm and a DA of 3% at the working point of (116.16, 41.12). The effective DA and the corresponding frequency map are shown in Fig. 5, and the other main parameters of this mode (denoted as mode 1) are summed up in Table 1.

In addition to find a larger DA and MA for the HEPS design with low beta functions at the ID section, another special interest is to achieve a dedicated commissioning mode with a larger ring acceptance. To this end, we repeat the above optimization process, however, with a released constraint on the beta functions at the long straight section, 1.5 m < $\beta_y$ < 40 m and 1.5 m < $\beta_x$ < 40 m. In this case the effective vertical DA can be further increased to ~4 mm with a working point of (115.31, 31.10). The main parameters of this mode (denoted as mode 2) are also listed in Table 1.

From the above analysis and the comparison of the parameters between these two modes and the baseline design in Table 1, one can learn that with the same lattice structure, it is somewhat difficult to significantly reduce the emittance; however, it is feasible to achieve a much better ring performance by re-matching the linear optics to realize weaker chromatic sextupoles and an optimal working point.

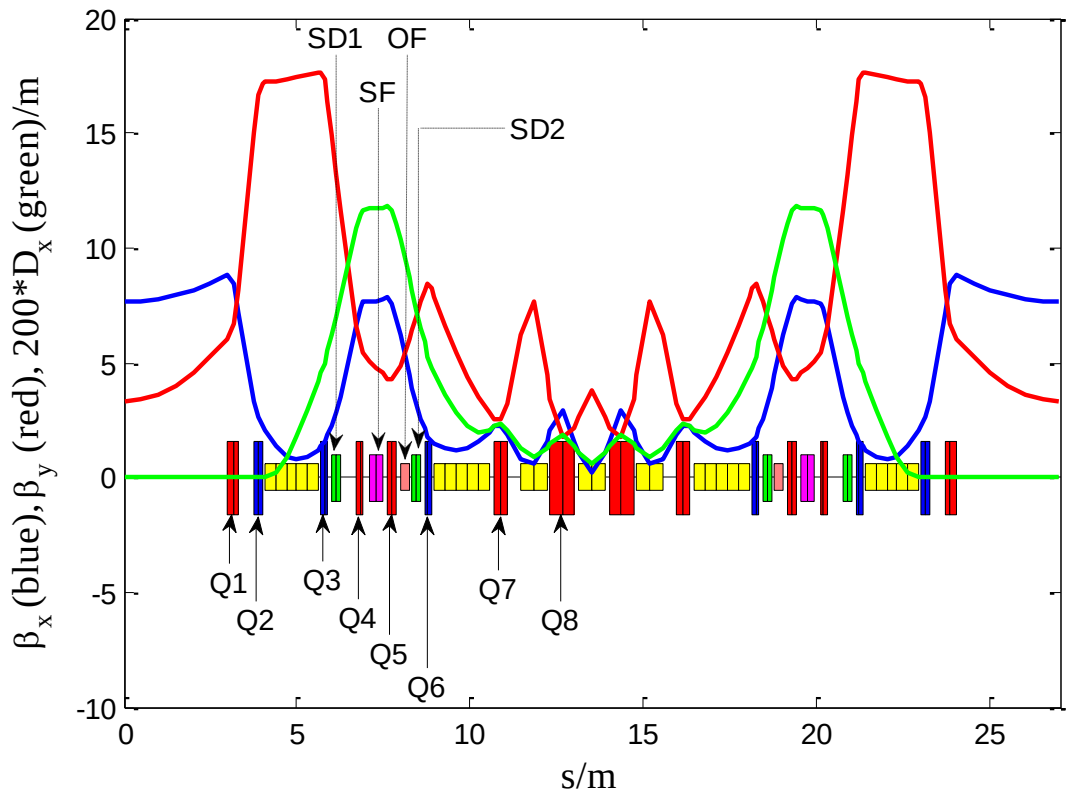

Fig. 1. (color online) The optical functions and the layout of the hybrid 7BA for the HEPS baseline design, with a mirror symmetry. The names of the quadrupoles and multipoles are marked in the plot.

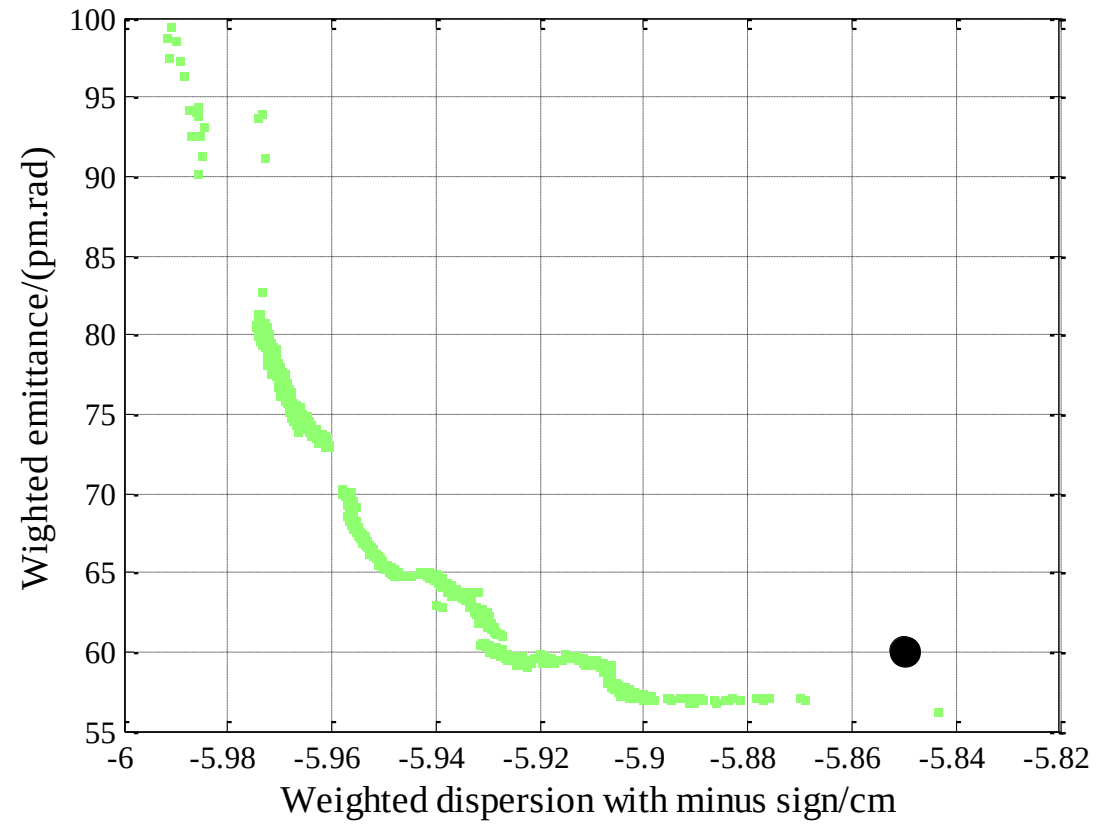

Fig. 2. Results of the 120th generation in the objective space, obtained with the genetic optimization based on the layout of the HEPS baseline design. The values of the objectives corresponding to the baseline design are also plotted as a black dot.

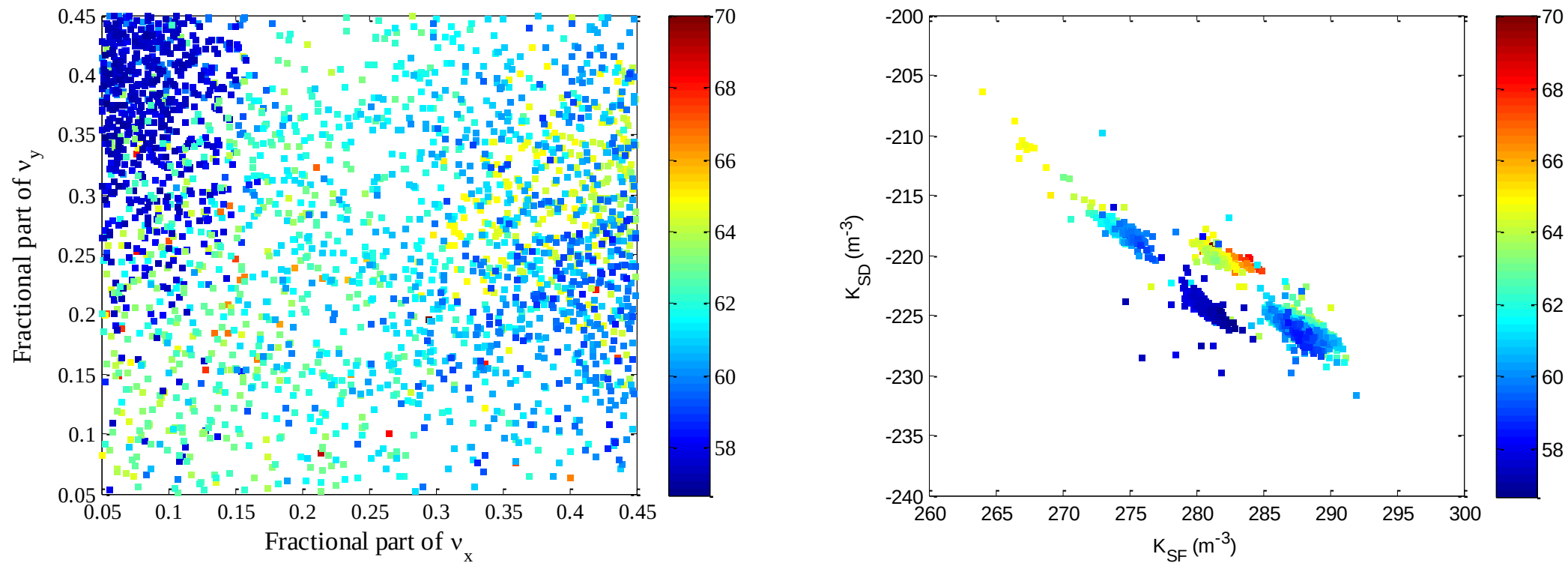

Fig. 3. (color online) Distribution of the results in the tune space (left) and the sextupole strength space (right), with the colors from blue to red, represent the emittances from lower to larger values.

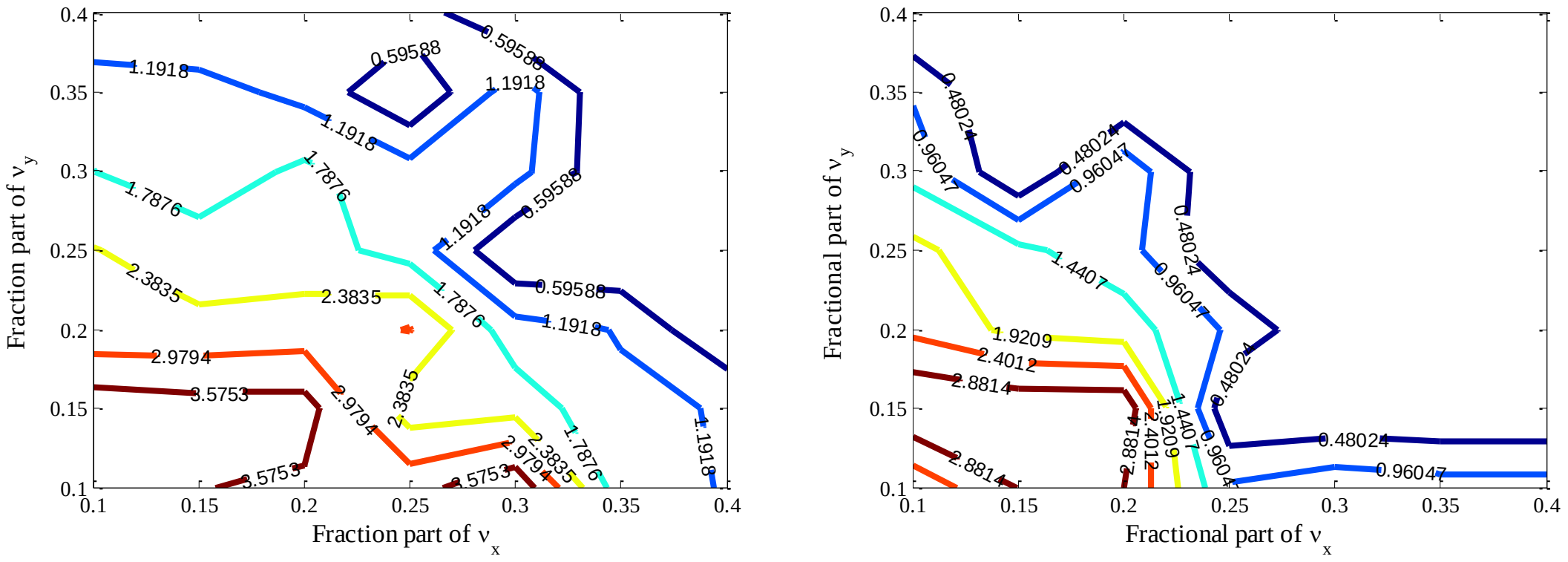

Fig.4. Contour plots of the effective vertical DA in the tune space, with MA larger than 2.5% (left) and 3% (right), respectively.

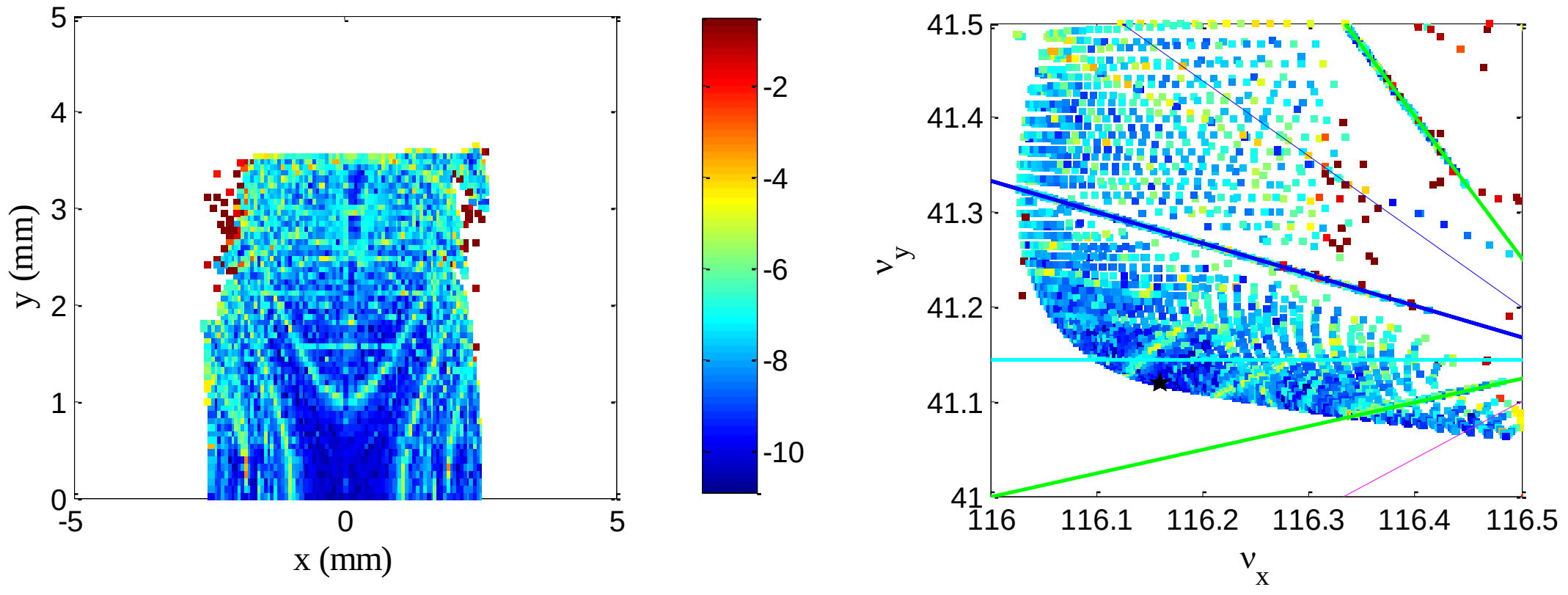

Fig. 5. (color online) The effective DA and the corresponding frequency map for the HEPS design, with the working point matched to (116.16, 41.12). The colors, from blue to red, represent the stabilities of the particle motion, from stable to unstable.

Table 1. Main parameters of the HEPS baseline design and two optimized modes

| | **Baseline design** | **Mode 1 (low beta)** | **Mode 2 (large beta)** |
|---|---|---|---|
| Working point (H/V) | 113.20/41.28 | 116.16/41.12 | 115.31/31.10 |
| Natural chromaticity (H/V) | –149/–128 | –214/–133 | –205/–168 |
| Beta functions in ID straight section (H/V)/m | 7.6/3.3 | 9/3.2 | 9/12.4 |
| Natural emittance $\varepsilon_0$/(pm·rad) | 60 | 59.4 | 58.6 |
| Lengths of (SD1/SD2/SF/OF)/m | 0.25/0.25/0.34/0.26 | 0.25/0.25/0.34/0.26 | 0.25/0.25/0.34/0.26 |
| Gradients of (SD1/SD2/SF)/$m^{-3}$ | –180/–240/203 | –208/–166/171 | –202/–136/169 |
| Gradient of OF/$m^{-4}$ | 25000 | 11700 | 13200 |
| Momentum acceptance | 2.4% | 3% | 3% |
| Vertical efficient DA size/mm | 2.2 | 3.5 | 4 |

## 3 Conclusions

In the design of a DLSR which typically has extremely strong nonlinearities, like HEPS, it is a great challenge to achieve an ultralow emittance and to well control the nonlinear aberrations simultaneously. In this study, keeping the lattice structure the same, we perform a global scan of the quadrupole strengths with an evolutionary genetic method, NSGA-II, to find candidate designs with different linear optical parameters, and then investigate the dependence of the nonlinear performance on specific optical parameters, e.g., the tunes. This approach proves to be as effective as the previous genetic optimizations using directly the DA as objective, but is more efficient, in optimization of the ring performance

Now the HEPS is in the design phase and the optimization of the HEPS design remains an open question. Also, the application of genetic optimization in the HEPS can be modified or extended. For instances, one can use the chromatic sextupole strengths which are more closely related to the nonlinear performance, rather than the dispersion at the sextupole as the optimization objective, or knob more element parameters in the optimization to search for an alternative layout that predicts a better ring performance. Actually such a study is underway, which will be reported in a forthcoming paper.